\documentclass{article}
\usepackage[T1]{fontenc}
\usepackage{wrapfig}
\usepackage[portuges,english]{babel}
\usepackage{amssymb,latexsym,amsmath,color,mathrsfs,ifsym,graphics,stmaryrd} 
\usepackage[colorlinks,linkcolor=blue,urlcolor=blue,citecolor=black,
plainpages=false,pdfpagelabels,breaklinks]{hyperref}

\title{Causality and the Modeling of the \\Measurement Process in Quantum Theory}
\author{{\sc Christian de Ronde}\thanks{Fellow Researcher of the Consejo
Nacional de Investigaciones Cient\'{\i}ficas y T\'ecnicas and Adjoint Professor of the National University Arturo Jauretche.}}
\date{\begin{center}
\begin{small} 
CONICET, Buenos Aires University - Argentina \\
Center Leo Apostel and Foundations of  the Exact Sciences\\
Brussels Free University - Belgium \\
\end{small}
\end{center}}

\usepackage[margin=3.5cm]{geometry}

\begin{document}
\maketitle

\begin{abstract}
\noindent In this paper we provide a general account of the causal models which attempt to provide a solution to the famous measurement problem of Quantum Mechanics (QM). We will argue that ---leaving aside instrumentalism which restricts the physical meaning of QM to the algorithmic prediction of measurement outcomes--- the many interpretations which can be found in the literature can be distinguished through the way they model the measurement process, either in terms of the {\it efficient cause} or in terms of the {\it final cause}. We will discuss and analyze why both, `final cause' and `efficient cause' models, face severe difficulties to solve the measurement problem. In contradistinction to these schemes we will present a new model based on the {\it immanent cause} which, we will argue, provides an intuitive understanding of the measurement process in QM. 
\medskip
\end{abstract}

\textbf{Keywords}: causality, models, explanation, measurement problem, quantum mechanics.

\renewenvironment{enumerate}{\begin{list}{}{\rm \labelwidth 0mm
\leftmargin 0mm}} {\end{list}}

\newcommand{\ita}{\textit}
\newcommand{\mcal}{\mathcal}
\newcommand{\mfrak}{\mathfrak}
\newcommand{\mbb}{\mathbb}
\newcommand{\mrm}{\mathrm}
\newcommand{\msf}{\mathsf}
\newcommand{\mscr}{\mathscr}
\newcommand{\lra}{\leftrightarrow}
\renewenvironment{enumerate}{\begin{list}{}{\rm \labelwidth 0mm
\leftmargin 5mm}} {\end{list}}

\newtheorem{dfn}{\sc{Definition}}[section]
\newtheorem{thm}{\sc{Theorem}}[section]
\newtheorem{lem}{\sc{Lemma}}[section]
\newtheorem{cor}[thm]{\sc{Corollary}}
\newcommand{\Proof}{\textit{Proof:} \,}
\newcommand{\cqd}{{\rule{.70ex}{2ex}} \medskip}

\bigskip

\bigskip

\bigskip

\bigskip

\section*{Introduction}

In this paper  we attempt to analyze and discuss the infamous measurement problem of QM. The solution to this problem has been discussed by many interpretations; however, as different as they might seem at first sight, all interpretations of QM found in the literature can be characterized in terms of two general types of causal explanation. The first is grounded on the {\it final cause} and implies a return of physics to the Aristotelian hylomorphic scheme configured in terms of the potential and actual realms. The second explanation to the quantum process of measurement has been given in terms of the {\it efficient cause} and implies an attempt to understand QM as related to the classical Newtonian representation of physics in terms of evolving actual properties. In this paper we attempt to put forward a new model of causal explanation based on the Spinozist notion of {\it immanent cause}. We will argue that our model provides an intuitive understanding of the measurement process in QM avoiding the many difficulties already found within the orthodox Aristotelian `final cause (hylomorphic) model' and the Newtonian `efficient cause model'. 
 
The paper is organized as follows. In section 1, we provide a general account of Aristotelian hylomorphism which relates, through the final cause, the potential and actual realms. It should be clear that we do not attempt to provide a complete nor detailed introduction to Aristotle's metaphysics, rather we intend to introduce the reader with some specific notions which will be used during the rest of the article. In section 2, we discuss how Newtonian mechanics eliminated the realm of potentiality and conceived actuality as the only relevant mode of existence for physical description. In this case it is only the efficient cause which accounts for the evolution of properties. In section 3 we discuss the historical background surrounding the ``quantum jumps'' and the difficulties it implies for physical representation. Section 4 analyzes the meaning of the projection postulate and its physical interpretation in terms of ``collapse'' and ``non-collapse'' models, making special emphasis on the notions of causality that each of these models implies. Section 5 presents an original solution to the relation between quantum superpositions and measurement outcomes through the introduction of the Spinozist {\it immanent cause}. Finally, in section 6, we present the conclusions of the article.

\section{Hylomorfism and the Final Cause: From Potentiality to Actuality}

The debate in Pre-Soctratic philosophy is traditionally understood as the contraposition of the Heraclitean and the Eleatic schools of thought. Heraclitus was considered as defending a theory of flux, a doctrine of permanent motion, change and instability in the world. This doctrine precluded, as both Plato and Aristotle stressed repeatedly, the possibility to develop certain knowledge about the world. In contraposition to the Heraclitean school we find Parmenides as the founder of the Eleatic school. Parmenides, as interpreted also by Plato and Aristotle, taught the non-existence of motion and change in reality, reality being absolutely One, and being absolutely Being. In his famous poem Parmenides stated maybe the earliest intuitive exposition of the principle of non-contradiction; i.e. that which {\it is} can only {\it be}, that which {\it is not, cannot be}. The problem of movement as presented by both Plato and Aristotle confronted these two metaphysical schemes. 

In order to solve the problem of movement, Aristotle developed a metaphysical system in which, through the notions of {\it actuality} and {\it potentiality}, he was able to articulate both the Heraclitean and the Eleatic schools of thought. On the one hand, potentiality defined an undetermined, contradictory and non-individual realm of existence, on the other, the mode of being of actuality was determined through the logical and ontological Principle of Existence (PE), Principle of Non-Contradiction (PNC) and Principle of Identity (PI). The notion of `entity' was presented as being partly potential and partly actual (see for a detailed discussion \cite{VerelstCoecke}). Through this notion the hylomorphic Aristotelian scheme was capable of unifying, of totalizing in terms of a ``sameness'', creating certain stability for knowledge to be possible. This representation or transcendent description of the world is considered by many the origin of metaphysical thought itself. Actuality was then understood as representing a mode of existence independent of observation. This is the way through which metaphysical thought was able to go beyond the  {\it hic et nunc}, creating a world beyond the world, a world represented through metaphysical concepts.

In the book $\Theta$ of {\it Metaphysics}, Aristotle [1046b5-1046b24] remarks there are two types of potentiality: ``[...] some potentialities will be non-rational and some will be accompanied by reason.'' For obvious reasons Aristotle calls these two potentialities `rational' and  `irrational'. Irrational potentiality implies a realm of `indefiniteness', a realm of `incompleteness' and `lack'. It is only when turning into actuality, that the potential is fulfilled, completed (e.g. the child becoming a man, the seed transforming into a tree) [1047b3-1047b14]. The path from irrational potentiality into actualization may be related to the {\it process} through which {\it matter} turns into {\it form}. The matter of a substance being the stuff it is composed of; the form, the way that stuff is put together so that the whole it constitutes can perform its characteristic functions. Through this passage substance becomes more perfect and, in this way, closer to God, {\it pure acto} [1051a4-1051a17].\footnote{As noticed by Verelst and Coecke \cite[p. 168]{VerelstCoecke}: ``change and motion are intrinsically not provided for in this [Aristotelian logical] framework; therefore the ontology underlying the logical system of knowledge is essentially static, and requires the introduction of a First Mover with a proper ontological status beyond the phenomena for whose change and motion he must account for.'' This first mover is God, {\it pure acto}, pure definiteness and form without the contradiction and evil present in the potential matter.} Because of this it makes no sense to consider the realm of irrational potentiality independently of actuality. It is the final cause which provided the link that closed the gap between both realms. 

Almost forgotten in the literature, which discusses the notion of potentiality exclusively in terms of irrational potentiality,\footnote{A good example of this state of affairs is given by the article of the {\it Stanford Encyclopedia of Philosophy} entry about Aristotle's metaphysics. When making reference to actuality and potentiality, the article only refers to the notion of  `potentiality' without any specific remark regarding the distinction put forward by Aristotle between rational and irrational potentiality.} the notion of {\it rational potentiality} is characterized by Aristotle as related to the problem of possessing a capability, a faculty [1046b5-1046b24], to what I mean when I say: ``I can'', ``I cannot''. As explicitly noticed by Aristotle, potentiality implies a mode of existence which must be considered as real as actuality. In chapter 3 of book  $\Theta$ of {\it Metaphysics} Aristotle introduces the notion of rational potentiality as independent of the actual realm.\footnote{Aristotle goes against the Megarians who considered actuality as the only mode of existence: ``There are some who say, as the Megaric school does, that a thing can act only when it is acting, and when it is not acting it cannot act, e.g. he who is not building cannot build, but only he who is building, when he is building; and so in all other cases. It is not hard to see the absurdities that attend this view. For it is clear that on this view a man will not be a builder unless he is building (for to be a builder is to be able to build), and so with the other arts. If, then, it is impossible to have such arts if one has not at some time learnt and acquired them, and it is then impossible not to have them if one has not sometime lost them (either by forgetfulness or by some accident or by time; for it cannot be by the destruction of the object itself, for that lasts forever), a man will not have the art when he has ceased to use it, and yet he may immediately build again; how then will he have got the art? [...] evidently potentiality and actuality are different; but these views make potentiality and actuality the same, so that it is no small thing they are seeking to annihilate. [...] Therefore it is possible that a thing may be capable of being and not be, and capable of not being and yet be, and similarly with the other kinds of predicate; it may be capable of walking and yet not walk, or capable of not walking and yet walk.'' [1046b29 - 1047a10]} That which exists within rational potentiality is then characterized as being capable of both contrary effects.\footnote{``Since that which is capable is capable of something and at some time and in some way ---with all the other qualifications which must be present in the definition---, and since some things can work according to a rational formula and their potentialities involve a formula, while other things are non-rational and their potentialities are non-rational, and the former potentialities must be in a living thing, while the latter can be both in the living and in the lifeless; as regards potentialities of the latter kind, when the agent and the patient meet in the way appropriate to the
potentiality in question, the one must act and the other be acted
on, but with the former kind this is not necessary. For the
non-rational potentialities are all productive of one effect each,
but the rational produce contrary effects, so that they would
produce contrary effects at the same time; but this is impossible.
That which decides, then, must be something else; I mean by this,
desire or choice.'' [1048a1-1048a24]}
That also means that what exists in potentiality is capable of being and not-being
at one and the same time.\footnote{``Every potentiality is at one and the same time a potentiality for the opposite; for, while that which is not capable
of being present in a subject cannot be present, everything that is
capable of being may possibly not be actual. That, then, which is
capable of being may either be or not be; the same thing, then, is
capable both of being and of not being.'' [1050b7-1050b28]}
The contradiction of being and not-being present in rational
potentiality is only dissolved when we consider the actual realm, where only
one of the terms is effectuated. Contrary to the case of irrational
potentiality, where a teleological cause places the end in
actuality, rational potentiality might be interpreted as a realm independent of actuality (see e.g., \cite{Agamben99}).

Although Aristotle presented at first both actual and potential realms as ontologically equivalent, from chapter 6 of book $\Theta$, he seems to have changed his mind; placing actuality in the central axis of his architectonic he suddenly began to relegate potentiality to a mere supplementary role. ``We have distinguished the various senses of `prior', and it is clear that actuality is prior to potentiality. [...] For the action is the end, and the actuality is the action. Therefore even the word `actuality' is derived from `action', and points to the fulfillment.'' [1050a17-1050a23] Aristotle then continues to provide arguments in this line which show ``[t]hat the good actuality is better and more valuable than the good potentiality.'' [1051a4-1051a17] The choice of Aristotle to take irrational potentiality and the final cause as the basis of his metaphysics determined the fate of western thought.

\section{Classical Physics and the Efficient Cause: The End of the Potential Realm}

The importance of potentiality, which was a central element in Aristotle's scheme was erased with the advent of modern science during Early Moderinty. As we have seen above, it could be argued that the seed of this move was already present in the Aristotelian architectonic, whose focus was clearly placed in the actual realm. In relation to the development of physics, the focus and preeminence was also given to actuality. The XVII century division between {\it res cogitans} and {\it res extensa} played in this respect an important role separating very clearly the realms of actuality and potentiality. The realm of potentiality, as a different (ontological) mode of the being, was neglected becoming not more than mere (logical) {\it possibility}. 

The philosophy which was developed after Descartes kept {\it res cogitans} (thought) and {\it res extensa} (entities as acquired by the senses) as separated realms.\footnote{The fact that {\it res cogitans}, the soul, was related to the {\it indefinite} realm of potentiality can be found already in Aristotle's {\it De Anima}. That {\it res extensa}, i.e. the entities as characterized by the principles of logic, relate to the actual realm is something that would be developed by Newton's mechanics. Indeed, after Descartes Newtonian physics would become in the XVIII century a physics of actual entities.} As Heisenberg \cite[p. 73]{Heis58} makes the point: ``Descartes knew the undisputable necessity of the connection, but philosophy and natural science in the following period developed on the basis of the polarity between the `res cogitans' and the `res extensa', and natural science concentrated its interest on the `res extensa'. The influence of the Cartesian division on human thought in the following centuries can hardly be overestimated, but it is just this division which we have to criticize later from the development of physics in our time.'' This materialistic conception of science based itself on the main idea that extended things exist as being definite, that is, in the actual realm of existence. With modern science the actualist Megarian path was recovered and potentiality dismissed as a problematic and unwanted guest. The transformation from medieval to modern science coincides with the abolition of Aristotelian hylomorphic metaphysical scheme ---in terms of potentiality and actuality--- as the foundation of knowledge. However, the basic structure of Aristotelian logic still remained the basis for correct reasoning. As noted by Verelst and Coecke:

\begin{quotation}
\noindent {\small ``Dropping Aristotelian metaphysics, while at the same time continuing to use Aristotelian logic as an empty `reasoning apparatus' implies therefore losing the possibility to account for change and motion in whatever description of the world that is based on it. The fact that Aristotelian logic transformed during the twentieth century into different formal, axiomatic logical systems used in today's philosophy and science doesn't really matter, because the fundamental principle, and therefore the fundamental ontology, remained the same. This `emptied' logic actually contains an Eleatic ontology, that allows only for static descriptions of the world.'' 
\cite[p. 7]{VerelstCoecke}}
\end{quotation}

It was Isaac Newton who was able to translate into a closed mathematical formalism both, the ontological presuppositions present in Aristotelian (Eleatic) logic and the materialistic ideal of {\it res extensa} together with actuality as its mode of existence.\footnote{We remark that the history of how Newtonian physics was interpreted is of course very complicated. The image that we are sketching here applies to the eighteenth century French Newtonians like Laplace which is the interpretation that became orthodox. For a detailed analysis of the worldpicture that was developed from Newton's mechanics see \cite{Dijksterhuis86}.} Potentiality was then completely erased from the Newtonian picture. In this sense, classical mechanics might be regarded as a physics of pure actualities. Indeed, in Newtonian mechanics the representation of the state of a physical system is given by a point in phase space $\Gamma$ and the physical magnitudes are represented by real functions over $\Gamma$. These functions commute in between each others and can be interpreted as possessing definite values independently of measurement, i.e. each function can be interpreted as being actual. Obviously, the term actual refers here to {\it preexistence} (within the transcendent representation) and not to the observation {\it hic et nunc}. Every physical system may be described exclusively by means of its actual properties. The change of the system may be described by the change of its actual properties. Thus, potential or possible properties are considered only epistemically as the points to which the system might arrive ---depending in the initial conditions--- in a future instant of time. As Dieks makes the point:

\begin{quotation}
\noindent {\small ``In classical physics
the most fundamental description of a physical system (a point in
phase space) reflects only the actual, and nothing that is merely
possible. It is true that sometimes states involving probabilities
occur in classical physics: think of the probability distributions
$\rho$ in statistical mechanics. But the occurrence of possibilities
in such cases merely reflects our ignorance about what is actual.
The statistical states do not correspond to features of the actual
system (unlike the case of the quantum mechanical superpositions),
but quantify our lack of knowledge of those actual features.''
\cite[p. 124]{Dieks10}}
\end{quotation}

\noindent Classical mechanics tells us via the equation of motion how the state of the system moves along the curve determined by the initial conditions in $\Gamma$ and thus, as any mechanical property may be expressed in terms of $\Gamma$'s variables, how all of them evolve. Moreover, the structure in which actual properties may be organized is the (Boolean) algebra of classical logic. Newtonian mechanics had not only done away with free will but also with the {\it final cause} which governed the most important part of Aristotle's scheme. Instead, it was now the {\it efficient cause} which was capable of articulating the evolution of actualities. In each instant of time the world was constituted by definite valued properties, i.e. an {\it actual state of affairs}. This is why we might also say that classical mechanics is in fact the physics of {\it pure actuality}.

\section{A-Causal Quantum Jumps and the End of \\Representation}

The rise of QM in 1900 placed, since its origin, serious obstacles to maintain an account of physical reality in terms of an actual state of affairs. The quantum  principle introduced by Max Planck had shaken the very foundation of Newtonian physics itself. As noted by Bohr (\cite{Bohr34}, p. 53): ``[the essence of quantum theory] may be expressed in the so-called quantum postulate, which attributes to any atomic process an essential discontinuity, or rather individuality, completely foreign to the classical theories and symbolized by Plank's quantum of action.''  Discreetness precluded a description in terms of space and time while indeterminacy threatened to break down the classical physical explanation in terms of the {\it efficient cause}. Nevertheless such non-classical principles were used to develop further understanding within the theory. Einstein was one of the first to make use of the quantum principle explaining through it the photoelectric effect in 1905, Bohr also used it explicitly in his 1913 model of the atom. Even Heisenberg developed his matrix mechanics and derived the indeterminacy principle in the mid 20s taking as a standpoint Planck's quantum postulate. The development of quantum theory, based on such non-classical standpoints was clearly threatening the ideal of physics as a representation of the world in classical terms. Many, including Einstein, were certainly uncomfortable with this situation. 

A new hope to recover physical representation was born with Schr\"{o}dinger's wave mechanics which attempted to develop de Broglie's ideas of matter waves ---restoring a classical picture to the theory of quanta. Schr\"{o}dinger was hoping that through his wave equations quantum jumps would soon disappear and causality would be restored. But contrary to his expectations, Born's interpretation of Schr\"{o}dinger's quantum wave function ---in terms of a probability density--- made explicit that the quantum jumps had come to stay. Born \cite[p. 57]{WZ} argued in his paper there was no causal explanation regarding the jumps between the possible (the quantum wave function) and the actual realms (the measurement outcomes): ``One gets no answer to the question, `what is the state after the collision' but only to the question, `how probable is a specified outcome of the collision'.'' Some months later, in Copenhagen, after Bohr had almost convinced Schr\"odinger that the quantum jumps could not be removed from the theory, he is quoted to have said: ``Had I known that we were not going to get rid of this damned quantum jumping,  I never would have involved myself in this business!'' Indeed, Schr\"odinger understood very clearly the fact that the impossibility to produce a causal model for QM seemed also to preclude the possibility of an {\it anschaulich}\footnote{As remarked in \cite{HilgevoordUffink01}: ``the closest translation of the term anschaulich is `visualizable'. But, as in most languages, words that make reference to vision are not always intended literally. Seeing is widely used as a metaphor for understanding, especially for immediate understanding. Hence, anschaulich also means `intelligible' or `intuitive'.''} ---intuitive or intelligible--- representation of the theory. He considered this condition of {\it Anschaulichkeit} to be an essential requirement on any acceptable physical theory. Heisenberg \cite[p. 172]{Heis27} declared in this respect that: ``We believe we have gained {\it anschaulich} understanding of a physical theory, if in all simple cases, we can grasp the experimental consequences qualitatively and see that the theory does not lead to any contradictions.''

QM introduced a debate between those who wanted to continue having, in different ways, physics as a description of the world ---such as Einstein, Schr\"{o}dinger, de Broglie and Pauli--- and those who, for very different reasons ---such as Bohr, Born and Dirac\footnote{Heisenberg might be regarded as a highly pragmatic character who shifted from an epistemological position close to that of Bohr to some kind of Platonic realism about mathematical symmetries and structures.}--- did not mind losing the ideal of representation, provided the theory did the job of predicting, at least probabilistically, the measurement outcomes of a given classical experimental arrangement. The first stance had their attempts proposed, firstly, by Louis de Broglie in 1924 with his pilot wave theory, and secondly by Schr\"{o}dinger, at the beginning of 1927, when he put forward his wave mechanics. Both attempts were supported by Einstein. Anyhow, both of them had very difficult problems to overcome and match a descriptive coherent interpretation of phenomena. The second stance had stood on the critic put forward by Ernst Mach to the use of dogmatic metaphysical concepts ---e.g., Newtonian absolute space and time. Instead, Mach had called for a recovery of a sensualistic science based on observability alone. Certainly, the idea that observable magnitudes had to be defined by the theory itself ---rather than consider them as self evident givens---, was the guide that Heisenberg had used in order to arrive to his matrix mechanics ---later on developed by Max Born, Pascual Jordan and himself in 1926. But matrix mechanics was too abstract for the physicists of the time, accustomed to work with differential equations and visualizable models \cite{QTen}. Born ---a mathematician himself--- was happy to have found a closed consistent mathematical formalism, that was enough for him. But it was not enough for Einstein whose research as a physicist was based on a theory that would be able to represent physical reality. At the end of 1927 Niels Bohr had come up with an explanation of his own, based on his notion of complementarity, which focused in fulfilling the consistency requirements of the quantum formalism in order to apply to the well known classical descriptions in terms of `waves' and `particles' \cite{BokulichBokulich}. 

According to Bohr \cite[p. 7]{WZ}: ``[...] the unambiguous interpretation of any measurement must be essentially framed in terms of classical physical theories, and we may say that in this sense the language of Newton and Maxwell will remain the language of physicists for all time.'' Bohr had found a new {\it a priori}: classical language ---which would serve to secure intersubjectivity. But, in order to close the circle, no ``new language'' was allowed from now on to enter the scene: ``it would be a misconception to believe that the difficulties of the atomic theory may be evaded by eventually replacing the concepts of classical physics by new conceptual forms.'' \cite[p. 7]{WZ} Heisenberg's uncertainty relations ---understood epistemically by Bohr \cite{HilgevoordUffink01}--- would then secure the knowledge provided by the more general principle of complementarity. Bohr had regained objectivity by watching quantum theory from a distance, standing on the well known heights of classical language. However, the position of Bohr presented a very unclear relation between the classical world and the quantum formalism, which, according to Bohr, did not seem to have a place in the classical conception of the world, but nevertheless, talked about it in terms of measurement outcomes. 

Quite independently of the many problems which remained for a coherent interpretation, the story was told and repeated once and again, that losing {\it representation} was not so bad after all. As it is clearly stated by Arthur Fine, this war was won by Bohr:

\begin{quotation}
\noindent {\small ``In the body of the paper [from 1925], Heisenberg not only rejects any reference to un-observables; he also moves away from the very idea that one should try to form any picture of a reality underlying his mechanics. To be sure, Schr\"{o}dinger, the second father of quantum theory, seems originally to have had a vague picture of an underlying wavelike reality for his own equation. But he was quick to see the difficulties here and, just as quickly, although reluctantly, abandoned the attempt to interpolate any reference to reality. These instrumentalist moves, away from a realist construal of the emerging quantum theory, were given particular force by Bohr's so-called `philosophy of complementarity'; and this non-realist position was consolidated at the time of the famous Solvay conference, in October of 1927, and is firmly in place today. Such quantum non-realism is part of what every graduate physicist learns and practices. It is the conceptual backdrop to all the brilliant successes in atomic, nuclear, and particle physics over the past fifty years. Physicists have learned to think about their theory in a highly non-realist way, and doing just that has brought about the most marvelous predictive success in the history of science.''
\cite[p. 88]{Fine86}}
\end{quotation}

\noindent Today, the so called ``Copenhagen Interpretation of QM''\footnote{The idea of a common interpretation of the founding fathers is difficult to maintain. Don Howard's research has shed some light to the creation of what he calls the ``Copenhagen Myth'' \cite{Howard04}. As remarked by Jones, the interpretation and development of QM ``was never a team effort. Sometimes, two or three would collaborate for a while, but mostly they were rivals who wanted their particular version of the new science to prevail. They had little enough in common.'' \cite[p. 10]{QTen}. See also \cite{Jammer66}.} is taught in Universities all around the world. Put in a nutshell this ``interpretation'' tells us how, following a set of rules,\footnote{See for example the orthodox texts: \cite{Cohen, Sakurai}.} one can calculate the prediction of probabilistic outcomes within any given experimental arrangement. The fifth postulate is the famously well known {\bf Projection Postulate}\footnote{Discussed in depth in the famous book by von Neumann \cite{VN}.} ({\bf PP}) which makes explicit ---as a reminder of the still unsolved problems--- those ``damned quantum jumps!''

\section{Modeling the Quantum Measurement Process: \\To Collapse or Not to Collapse?}

Classical texts that describe QM axiomatically begin stating that
the mathematical interpretation of a quantum system is a Hilbert
space, that pure states are represented by rays in this space,
physical magnitudes by self-adjoint operators on the state space and
that the evolution of the system is ruled by the Schr\"{o}dinger
equation. Possible results of a given magnitude are the eigenvalues
of the corresponding operator obtained with probabilities given 
by the Born rule. In general, the state previous to the
measurement is a linear superposition of eigenstates corresponding
to different eigenvalues of the measured observable. This gives rise to the infamous measurement problem.\\

\noindent {\it {\bf Measurement Problem:} Given a specific basis (or context), QM describes mathematically a quantum state in terms of a superposition of, in general, multiple states. Since the evolution described by QM allows us to predict that the quantum system will get entangled with the apparatus and thus its pointer positions will also become a superposition,\footnote{Given a quantum system represented by a superposition of more than one term, $\sum c_i | \alpha_i \rangle$, when in contact with an apparatus ready to measure, $|R_0 \rangle$, QM predicts that system and apparatus will become ``entangled'' in such a way that the final `system + apparatus' will be described by  $\sum c_i | \alpha_i \rangle  |R_i \rangle$. Thus, as a consequence of the quantum evolution, the pointers have also become ---like the original quantum system--- a superposition of pointers $\sum c_i |R_i \rangle$. This is why the measurement problem can be stated as a problem only in the case the original quantum state is described by a superposition of more than one term.} the question is why do we observe a single outcome instead of a superposition of them?}\\

Due to the existence of quantum superpositions, in order to give an account of the state of the system after the appearance of a particular result, one needs to add the {\bf PP}.  In von Neunmann's \cite[p. 214]{VN} words: ``Therefore, if the system is initially found in a state in which the values of $\mathcal{R}$ cannot be predicted with certainty, then this state is transformed by a measurement $M$ of $\mathcal{R}$ into
another state: namely, into one in which the value of $\mathcal{R}$
is uniquely determined. Moreover, the new state, in which $M$ places
the system, depends not only on the arrangement of $M$, but also on
the result of $M$ (which could not be predicted causally in the
original state) ---because the value of $\mathcal{R}$ in the new
state must actually be equal to this $M$-result''.\footnote{Or in
Dirac's words: ``When we measure a real dynamical variable $\xi$,
the disturbance involved in the act of measurement causes a jump in
the state of the dynamical system. From physical continuity, if we
make a second measurement of the same dynamical variable $\xi$
immediately after the first, the result of the second measurement
must be the same as that of the first.'' \cite[p. 36]{Dirac74}} However, the {\bf PP} does not provide a solution to the measurement problem by itself. The solution requires an explanation which relates in a causal model a superposition of multiple states and the one measurement outcome.  
 
It should be remarked that the measurement problem presupposes that the basis (or context) ---directly related to a measurement set up--- has been already determined (or fixed). Thus it should be clear that there is no question regarding the contextual character of the theory within this specific problem. As we have argued extensively in \cite{deRonde16b}, the measurement problem has nothing to do with contextuality. The measurement problem raises when, within a definite basis, the actualization process is considered. There is then a mix of subjective and objective elements when the recording of the experiment takes place ---as Wigner clearly explained in what became to be known as the `Wigner's friend {\it Gedenakenexperiment}' \cite{WZ}. The problem here is the coherency between the physical representation provided when the measurement was not yet performed, and the system is described in terms of a quantum superposition; and when we claim that ``we have observed a single measurement outcome'', which is not described by the theory. Since there is no physical representation of ``the collapse'', the subject (or his friend) seems to define it explicitly. The mixture of objective and subjective is due to an incomplete description of the state of affairs within the quantum measurement process (or ``collapse''). 

There are different ways to provide a physical account of the projection postulate and attempt to solve the measurement problem. However, all of them can be considered in terms of two causal models of the measurement process. The first explanation is provided via a collapse-model which makes use of the final cause; the second one is presented in terms of non-collapse models which attempt to restore the efficient cause as the key notion to account for the evolution of quantum properties. In the following, we consider both collapse and non-collapse models in some detail.

\subsection{Collapse Models and the Final Cause}

Collapse interpretations have their most important proponent in orthodoxy, a pseudo-instrumentalist perspective which goes ---unfortunately--- many times by the name of ``the Copenhagen interpretation of QM''. This interpretation, proposed by Heisenberg in his book, {\it Physics and Philosophy} \cite{Heis58}, presented a ``collapse'' model in order to explain the process of measurement in QM. According to it, the stochastic ``jump'' takes place from the state previous to the measurement to the eigenstate corresponding to the measured eigenvalue. As explained by Heisenberg:

\begin{quotation}
\noindent {\small ``The observation itself changes the probability function discontinuously; it selects of all possible events the actual one that has taken place. Since through the observation our knowledge of the system has changed discontinuously, its mathematical representation also has undergone the discontinuous change and we speak of a `quantum jump.' When the old adage `Natura non facit saltus' is used as a basis for criticism of quantum theory, we can reply that certainly our knowledge can change suddenly and that this fact justifies the use of the term `quantum jump.'

Therefore, the transition from the `possible' to the `actual' takes place during the act of observation. If we want to describe what happens in an atomic event, we have to realize that the word `happens' can apply only to the observation, not to the state of affairs between two observations. It applies to the physical, not the psychical act of observation, and we may say that the transition from the `possible' to the `actual' takes place as soon as the interaction of the object with the measuring device, and thereby with the rest of the world, has come into play; it is not connected with the act of registration of the result by the mind of the observer. The discontinuous change in the probability function, however, takes place with the act of registration, because it is the discontinuous change of our knowledge in the instant of registration that has its image in the discontinuous change of the probability function.''
\cite[p. 54]{Heis58}}
\end{quotation}

\noindent In order to make sense of the jump, Heisenberg recovered potentiality for physics going back to the hylomorphic Aristotelian scheme (section 1). However, as it became soon clear, this solution also introduces a problematic subjective element within the description itself. As we already mentioned, this model has deep problems concerning the meaning of objective physical reality in QM. Indeed, the intromission of an observer seems to destroy the possibility of an objective representation. Another deep problem of this scheme is that the realm of potentiality is only defined in terms of its actualization. As a consequence, the quantum potential existents remain in an ontological limbo, not truly real but not truly unreal. 

\begin{quotation}
\noindent {\small ``In the experiments about atomic events we have to do with things and facts, with phenomena that are just as real as any phenomena in daily life. {\it But the atoms or the elementary particles themselves are not as real; they form a world of potentialities or possibilities rather than one of things or facts.}'' \cite[p. 160]{Heis58} (emphasis added)}
\end{quotation}

Continuing the line of work of Heisenberg's hylomorphic interpretation in the new physics, Constantin Piron has been one of the leading figures in developing the notion of potentiality within the logical structure of QM \cite{Piron76, Piron81, Piron83}. The Geneva approach to quantum logic attempts to consider quantum physics as related to the realms of actuality and potentiality analogously to classical physics. According to the Geneva school, both in classical and quantum physics measurements will provoke
fundamental changes of the state of the system. What is special for
a classical system, is that `observables' can be described by
functions on the state space. This is the main reason that a
measurement corresponding to such an observable can be left out of
the description of the theory `in case one is not interested in the
change of state provoked by the measurement', but `only interested
in the values of the observables'. It is in this respect that the
situation is very different for a quantum system. Observables can
also be described, as projection valued measures on the Hilbert
space, but `no definite values can be attributed to such a specific
observable for a substantial part of the states of the system'. For
a quantum system, contrary to a classical system, it is not true
that `either a property or its negation is actual'. A physical property, never mind whether a classical or quantum one, is specified as what corresponds to a set of definite experimental projects \cite{Smets05}. A {\it Definite Experimental Project} (DEP) is an experimental procedure (in fact, an equivalence
class of experimental procedures) consisting in a list of actions
and a rule that specifies in advance what has to be considered as a
{\it positive} result, in correspondence with the {\it yes} answer
to a dichotomic question. Each DEP tests a property. A given DEP is
called {\it certain} (correspondingly, a dichotomic question is
called {\it true}) if it is sure that the positive response would be
obtained when the experiment is performed or, more precisely, in
case that whenever the system is placed in a measurement situation
then it produces certain definite phenomenon to happen. A physical
property is called {\it actual} in case the DEPs which test it are
certain and it is called {\it potential} otherwise. Whether a
property is actual or potential depends on the state in which one
considers the system to be. 

Though in this approach both actuality and potentiality are considered as modes of being, actual properties are considered as attributes that {\it exist}, as  elements of (EPR) physical reality, while potential properties are not conceived as existing in the same way as real ones. They are thought as {\it
possibilities} with respect to actualization, because potential
properties may be actualized due to some change in the state of the
system. In this case the superposition provides a measure ---given
by the real numbers which appear in the same term as the state---
over the irrational potential properties which could become actual
in a given situation. Thus, potentiality, as in the classical
physical sense, can be regarded as {\it irrational potentiality}, as
referring to a future in which a given property can become actual.

Also closely related to the development of Heisenberg in terms of
potentialities stands the development of Margenau and Popper in
terms of latencies, propensities or dispositions. As recalled by
Su\'arez \cite{Suarez07}, Margenau was the first to introduce in
1954 a dispositional idea in terms of what he called {\it
latencies}. In Margenau's interpretation the probabilities are given
an objective reading and understood as describing tendencies of
latent observables to take on different values in different contexts
\cite{Margenau54}. Later, Karl Popper \cite{Popper82}, followed by
Nicholas Maxwell \cite{Maxwell88}, proposed a propensity
interpretation of probability. Quantum reality was then
characterized by irreducibly probabilistic real propensity
(propensity waves or propensitons).\footnote{The realist position of
Popper seemed in this respect much more radical than the
interpretation of Heisneberg in terms of potentia. Heisenberg
\cite[pp. 67-9]{Heis58} seemed to remain within a subjectivist
definition of such potentia: ``Such a probability function [i.e. the
statistical algorithm of quantum theory] combines objective and
subjective elements. It contains statements on possibilities, or
better tendencies (`potentiae' in Aristotelian philosophy), and such
statements are completely objective, they don't depend on any
observer the passage from the `possible' to the real takes place
during the act of observation''. This was something Popper was
clearly against.} More recently, Mauricio Su\'arez has put forward a
new interpretation in which the quantum propensity is
intrinsic to the quantum system and it is only the manifestation of
the property that depends on the context \cite{Suarez04b,
Suarez07}. Mauro Dorato has also advanced a dispositional approach
towards the GRW theory \cite{Dorato06, Dorato11}. But the GRW
theory ---named after their creators: Ghirardi, Rimmini and Weber
\cite{GRW}--- is a dynamical collapse model of non-relativistic
QM which modifies the linearity of Schr\"odinger's
equation and thus, it does not respect the orthodox quantum formalism. 

The general solution to the measurement problem provided by final cause models face serious difficulties. In the first place, there is no observation of any real collapse taking place in the measurement process. As Dieks points out: 

\begin{quotation}
\noindent {\small ``Collapses constitute a process of evolution that conflicts with the evolution governed by the Schr\"{o}dinger equation. And this raises the question of exactly when during the measurement process such a collapse could
take place or, in other words, of when the Schr\"{o}dinger equation
is suspended. This question has become very urgent in the last
couple of decades, during which sophisticated experiments have
clearly demonstrated that in interaction processes on the
sub-microscopic, microscopic and mesoscopic scales collapses are
never encountered.'' \cite[p. 120]{Dieks10}}
\end{quotation}

\noindent In the second place, all these approaches lack a metaphysical or categorical definition of the potential realm they talk about. As remarked by Mauro Dorato with respect to propensity or dispositionalist type interpretations: 

\begin{quotation}
\noindent {\small ``[...] dispositions express, directly or indirectly, those regularities of the world around us that enable us to predict the future. Such a predictive function of dispositions should be attentively kept in mind when we will discuss the `dispositional nature' of microsystems before measurement, in particular when their states is not an eigenstate of the relevant observable. In a word, the use of the language of `dispositions' does not by itself point to a clear ontology underlying the observable phenomena, but, especially when the disposition is irreducible, refers to the predictive regularity that phenomena manifest. {\it Consequently, attributing physical systems irreducible dispositions, even if one were realist about them, may just result in more or less covert instrumentalism.}'' \cite[p. 4]{Dorato06} (emphasis added)}\end{quotation}

\noindent Dorato's criticism to dispositions can be also extended to all interpretations which introduce a potential realm and the final cause in order to ``solve'' the measurement problem and explain why we observe single measurement outcomes instead of superpositions. We remark that this criticism is a deathly strike to hylomorphic schemes which following a realist perspective attempt to provide an understanding of the formalism beyond the mere ---instrumentalist--- reference to measurement outcomes. These models fail unless they are capable of providing an explanation of the content of the potential or dispositional realm they introduce.

\subsection{Non-Collapse Models and the Efficient Cause}

Put in a nutshell, non-collpase interpretations attempt to ``restore a classical way of thinking about {\it what there is}.'' \cite[p. 74]{Bacciagaluppi96}  And what there is, is a definite valued set of properties constituting an {\it actual state of affairs}. There is one single realm of existence, actuality, in which the evolution of properties is governed by the efficient cause. Possibility ---as in classical physics--- is only epistemic, and reflects our ignorance about {\it what there is}. Non-collapse interpretations deny the existence of a ``collapse'' (physical interaction) during measurement and claim ---in different ways--- that the quantum superposition also describes ---at least partly--- the actual realm. 

Many worlds interpretation (MWI) of QM is one of such non-collapse interpretations which has become an important line of investigation within the foundations of quantum theory domain. It is considered to be a direct  conclusion from Everett's first proposal in terms of `relative states' \cite{Everett57}.\footnote{Jeffrey Barrett has confronted this reading of Everett as related to many worlds. See \cite{Barrett11, Barrett16}.} Everett's idea was ``to let QM find its own interpretation'', doing justice to the symmetries inherent in the Hilbert space formalism in a simple and convincing way \cite{DeWittGraham73}. The solution proposed to the measurement problem is provided by assuming that each one of the terms in the superposition is {\it actual} in its own correspondent world (or branch). Thus, it is not only the single value which we see in `our world' which gets actualized but rather, that a branching of worlds takes place in every measurement, giving rise to a multiplicity of worlds with their corresponding actual values. The possible splits of the worlds are determined by the laws of QM. In this case, there is no need to conceptually distinguish between possible and actual because each state is actual inside its own branch and the {\it eigenstate-eigenvalue link} is maintained in each world. As remarked by Everett \cite[pp. 146-7]{Everett73} himself: ``The whole issue of the transition from `possible' to `actual' is taken care of in the theory in a very simple way ---there is no such transition, nor is such a transition necessary for the theory to be in accord with our experience. From the viewpoint of the theory all elements of a superposition (all `branches') are `actual', none any more `real' than the rest.''

However, regardless of the fact many worlds does provide a picture of quantum superpositions in terms of many different worlds or branches, it is still not clear if they truly solve the measurement problem. In fact, the many questions concerning the process of ``collapse'' (of the superposed quantum state) are simply replaced by the questions regarding the process of ``branching'' (which takes place between multiple worlds). The original questions with respect to the `collapse process' are simply translated into questions regarding the `branching process'. Is the branch taking place independently of observation? How is the branching process modeled? When exactly does the branching take place? etc. Still today these questions are being debated in the literature and there is still no consensus if the many worlds interpretation is capable of answering them in a satisfactory manner (see e.g.,  \cite{DawinThebault15, Jansson16, Kastner14}).

Another well known non-collapse proposal is the so called modal interpretation (MI) of QM. This approach states that superpositions remain always intact, independent of the result of the actual observation.\footnote{Van Fraassen discusses the problems of the collapse of the quantum wave function in \cite{VF91}, section 7.3. See also \cite{Dickson98}. Dieks \cite[p. 182]{Dieks88a} argues that: ``[...] there is no need for the projection postulate. On the theoretical level the full superposition of states is always maintained, and the time evolution is unitary. One could say that the `projection' has been shifted from the level of the theoretical formalism to the semantics: it is only the empirical interpretation of the superposition that the component terms sometimes, and to some extent, receive an independent status.''} One might say that the {\it eigenstate-eigenvalue link} is here accepted only in one direction, implying that given a state that is an eigenstate there is a definite value of the corresponding magnitude, i.e. its eigenvalue, but not the other way around. ``In modal interpretations the state is not updated if a certain state of affairs becomes actual. The non-actualized possibilities are not removed from the description of a system and this state therefore codifies not only what is presently actual but also what was presently possible. These non-actualized possibilities can, as a consequence, in principle still affect the course of later events.'' \cite[p. 295]{Vermaas99a} There are thus, within MI, two realms or levels given by the possible and the actual.\footnote{These levels are explicitly formally accounted for in both van Fraasen and Dieks MI. While van Fraassen distinguishes between the `dynamical states' and the `value states', Dieks and Vermaas consider a distinction between `physical states' and `mathematical states' \cite{VermaasDieks95}.} The passage from the possible to the actual is given through different interpretational rules which change depending on the particular version  \cite{Vermaas99a}. 

Leaving aside van Fraassen's empiricist stance according to which: {\it modalities are in our theories, not in the world};\footnote{Dieks has taken a stance in favor of an humean position with respect to modality \cite{Dieks10}.} there are several realistic MIs in which the ideal of an actual state of affairs is restored and possibilities are considered in classical terms. As remarked by Dieks: ``the probabilities occurring in the modal interpretation have the same status as classical probabilities and have the usual classical interpretations.'' \cite[p. 15]{Dieks07} While Dieks original version assumes the existence of one single actual property, the Bohmian versions of Bub and Bacciagaluppi and Dickson assume the existence of a set of actual definite valued properties \cite{Bub92, BacciagaluppiDickson97}. In both cases possibility remains epistemic and only actuality is regarded as real. This is the reason why MI are called by Bacciagaluppi ``a hidden variable theory'' \cite{Bacciagaluppi96}. As explained by Dieks: ``In our search for definite-valued observables it is possible to include interpretations like the Bohm interpretation if we allow for the possibility that there is a preferred observable $R$ that is always definite, for all quantum states (in the Bohm theory position plays this role). The situation in which no privileged  observable exists then becomes a special case.'' \cite[p. 6]{Dieks07} Since the recovery of an actual state of affairs plays a major role within some of the realist versions of the MI the connection to many worlds interpretations might be also regarded as quite direct. According to Dieks: ``There is perfect equivalence [to the MI] in the sense that the many-worlds interpretation is defined by the condition that each element of the measure space corresponds to an actual states of affairs, whereas the probabilistic alternative is defined by the condition that each element may correspond to the one actual (but unspecified) state of affairs.'' \cite[p. 17]{Dieks07}

Unfortunately, the MIs attempt is to restore a ``Newtonian'' {\it efficient cause}-model of interacting actual (definite valued) properties, also faces serious problems. The first is that most properties within the orthodox formalism cannot be regarded as definite valued properties due to KS type theorems \cite{Bacciagaluppi95, RFD14, Vermaas97}. The second problem is that quantum superpositions considered within the measurement problem might be composed by contradictory properties or states such as `the atom is decayed' and `the atom is not decayed' (see \cite{daCostadeRonde13}). Even though these properties cannot be related to actual ones simultaneously, for that would violate PNC, both terms relate to possible outcomes. Thus, possible properties remain, just like propensities or dispositions, also in an ontological limbo regarding existence. None of them is truly real (actual), but all of them are still {\it possible}. The problem is how something which is not real can suddenly become real? Furthermore, as remarked by Dieks \cite[pp. 124-5]{Dieks10}, classical possibilities never interact: ``In classical physics the most fundamental description of a physical system (a point in phase space) reflects only the actual, and nothing that is merely possible. It is true that sometimes states involving probabilities occur in classical physics: think of the probability distributions $\rho$ in statistical mechanics. But the occurrence of possibilities in such cases merely reflects {\it our ignorance} about what is actual.'' So, if QM talks about classical possibilities, how to explain their interaction during their evolution and within the process of {\it entanglement}?\footnote{See for a detailed discussion \cite{deRonde16}.}

\section{A New Non-Collapse Model of Quantum Measurements in Terms of the Immanent Cause}

As we discussed above (section 1), due to his choice of a teleological scheme ---based on irrational potentiality and supplemented by the final cause--- Aristotle closed the door to a different development of potentiality in terms of an independent mode of existence to that of actuality. As remarked by Wolfgang Pauli: 

\begin{quotation}
\noindent {\small ``Aristotle [...] created the important concept of {\small {\it potential being}} and applied it to {\small {\it hyle}}. [...] This is where an important differentiation in scientific thinking came in. Aristotle's further statements on matter cannot really be applied in physics, and it seems to me that much of the confusion in Aristotle stems from the fact that being by far the less able thinker, he was completely overwhelmed by Plato. He was not able to fully carry out his intention to grasp the {\small \emph{potential}}, and his endeavors became bogged down in early stages.'' \cite[p. 93]{PauliJung}}
\end{quotation}

\noindent This choice of relegating the potential realm to the actual one, still resonates in both collapse and non-collapse models of the quantum measurement process discussed in the previous section. These models have concentrated their efforts in justifying observations in the actual realm. While collapse models have focused on the process of actualization, going back to Aristotle's hylomorphic scheme and the {\it final cause}, non-collapse models have tried to eliminate completely the potential realm, going back to a classical physical description in terms of actual properties evolving according to an {\it efficient cause}. As we have seen above, both final cause and efficient cause type models face serious difficulties and problems which have not been solved ---at least, up to the present--- in a satisfactory manner. In this last section we would like to propose a different model which investigates the possibilities of {\it rational potentiality} and its connection to the {\it immanent cause}.

Our proposal begins with the definition of a mode of existence completely independent of the actual realm. The definition of this realm, which we call ontological potentiality ---closely related to rational potentiality---, will allow us to produce a very different model of the quantum measurement process. According to our model, just like entities are existents within the actual realm, {\it immanent powers} with definite {\it potentia} are existents within the potential realm. It should be noticed that even though there is a vast literature which investigates causal powers, our approach differs radically from these ontological schemes. Our ontology presents a very different notion of power which has an existence independent of the actual realm relating to it only through an immanent form of causation. 

Given a quantum superposition, $\sum c_i | \alpha_i \rangle$, we interpret each ket, $| \alpha_i \rangle$, as an immanent power; and each term which accompanies a ket in square modulus, $|c_i|^2$, as its respective potentia. An immanent power can be thought as a capability which exists in a potential realm and can produce an actual effectuation (see for a detailed analysis postulates IV and V of the Appendix). However, immanent powers do not restrict themselves to such actualization process ---as causal powers do. An actual effectuation reflects only a partial expression (within the actual realm) of a power. While the actualization of causal powers is ruled by the {\it final cause} the actualization of immanent powers is ruled by the {\it immanent cause}. In this way we detach immanent powers from a teleological definition grounded in the actual realm.

The immanent cause goes back to the discussion ---present already in Medieval philosophy--- concerning God and its attributes. It was Spinoza who might have taken this notion to its most extreme expression in the {\it Ethics}. Spinoza's maxim, {\it Deus sive Natura} (see for example: \cite{Nadler13}), implies the idea that God is Nature, and that everything is an expression of God. In the {\it Short Treatise}, as remarked by Melamed \cite{Melamed}, Spinoza characterizes an immanent cause as one in which (1) the agent and the one acted on are not different, (2) the agent ``acts on himself,'' (3) effect ``is not outside itself,'' and (4) the effect is part of the cause. The immanent cause allows for the expression of effects remaining both in the effects and its cause. It does not only remain in itself in order to produce, but also, that which it produces stays within. Thus, in its production of actual effects the potential does not deteriorate by becoming actual ---as in the case of the hylomorphic scheme and causal powers (see section 1, p. 4 of this paper).\footnote{For a more detailed discussion of the notion of immanent cause we refer to \cite[Chapter 2]{Melamed}.}

Immanent powers produce, apart from actual effectuations, also {\it potential effectuations} which take place in potentiality and remain independent of what happens in the actual realm. Within our model, while potential effectuations describe the ontological interactions between immanent powers and their potentia, actual effectuations are only epistemic expressions of the potentia of powers. Quantum physical reality discusses about immanent powers with definite potentia which exist in the potential realm. Our ontology of immanent powers needs to be considered in QM on equal footing to the ontology of entities in classical physics.\footnote{This idea, that classical physics describes the actual realm while QM describes a potential realm has forced us to develop an ontic pluralist scheme which also confronts the orthodox reductionistic inter-theoretic view according to which there must necessarily exist a quantum to classical limit \cite{deRonde16c}. The discussion of which exceeds the space of this paper.} As we shall now see, our metaphysical scheme allows us to model the process of measurement within quantum theory in an intuitive manner. 

The immanent cause plays an essential role in our model of measurement, for it allows us to retain the existence of powers independently of a particular actual effectuation. Applied to the measurement problem we see how the immanent cause allows us to relate quantum superpositions to actual observations in a radically different way as it has been done up to the present ---through the final cause and the efficient cause. A superposition containing a set of powers can be now considered to remain always existent, independently of the expression of actual results which do not affect the set of powers themselves. Measurement outcomes become actual effectuations which only express the set of immanent powers ---formally represented by the kets within a quantum superposition--- in the actual mode of existence. However, regardless of their particular actualization, immanent powers (described by superpositions) remain evolving deterministically ---according to the Schr\"{o}dinger equation--- in the potential mode of existence, interacting with other different sets of powers (superpositions) through entanglement; each time producing new potential effectuations. Some of these aspects might reminds us of some main features of possibility within the modal interpretation itself, now read from an ontological perspective. An analogy can be useful in order to picture the physical model we attempt to put forward for the quantum measurement process. Imagine two baseball players called Matthias and John. Regarding the baseball game everyone can understand if I argue that both Matthias and John possess a definite set of potentia with respect to the powers of batting, running and pitching. What does it mean that, for example, Matthias possesses the power of batting with a potentia of 0.9? It obviously means  that he is a very good batter. That if I throw 100 balls to him, he will be capable to batt approximately 90 balls. This also means that if I would like to learn (at the epistemic level) about the (ontic) power of Matthias to batt I obviously need to do statistics. The more statistics I make the better I will learn about the potentia of his power to batt. Of course, exactly the same applies to John, if we would like to learn about his powers of batting, running and pitching, we would also need to perform a statistical analysis for each one of these powers in order to know their respective potentia. Indeed, the statistical data we can obtain from the performance of each player in many baseball games become in our model an (epistemic) measure of the (ontic) potentia of the powers in question. From this perspective, QM talks about the powers and their potentia, how they evolve and interact; and not about the particular observations by subjects of actual effectuations. 

The measurement problem now finds an intuitive explanation within our model. Through the introduction of the immanent cause it is possible to argue that the actual effectuation of a power within a quantum superposition is only an expression of its potentia at the epistemic level. The fact that Matthias can batt 0.9 of the times does not imply in any way that if I throw a ball to Matthias he will batt the ball. This possibility is completely indetermined since the notion of power is a  statistical notion quantitatively defined in terms of its specific potentia. If John can batt with a potentia of 0.5, this means that only half of the times he will be able to batt the ball. It is more probable that Matthias batts a ball than John, however, this is in no way determined before the actual effectuations take place. Exactly the same happens in a Stern-Gerlach experiment, now understood in terms of powers. The fact that the outcome is `+' or `-' is in general ---except for very specific cases--- probabilistic. Within our model this probability reflects an objective feature of reality, namely, the measure of a power in terms of its specific potentia \cite{deRonde16a}. 

In classical physics, as we argued above, the ontological level has been exclusively considered in terms of the actual realm. In this case actual observation collapses with the actual mode of existence. The single (actual) observation of a property (in the actual realm of existence) is enough to characterize the property completely. On the contrary, a power exiting in the potential realm, cannot be characterized through a single observation. To characterize completely the power $P_i$ we require a statistical measure which indicates its potentia $p_i$. The power possesses an {\it intensive existence} which, contrary to classical properties is not either 0 or 1, but a number pertaining to the closed interval $[0,1]$. Thus, to characterize one power we require a statistical measure of many actual effectuations. This is a strong impediment to equate observation with reality. Unlike the case of classical properties which are characterized in binary terms, an observation provides only a partial access to the power which requires many observations in order to grasp its potentia. 

Our understanding of the potential and actual realms within QM implies an inversion regarding the epistemic-ontic relationship imposed by classical physics. In classical physics, actuality has been always understood, since Newton, as characterizing reality itself. This has made possible to collapse and confuse the realist and empiricist meaning of actuality. On the one hand, the realist understanding of actuality as a mode of existence, and on the other hand, the empiricist understanding of actuality as a {\it hic et nunc} observation. Within classical physics, potentiality was then relegated to an epistemic level of analysis concerning the ignorance about the future (actual) state of affairs. Within our approach, QM has inverted the relation between on the one hand, potentiality and actuality, and on the other hand, ontology and epistemology. While in classical physics actuality is ontological and possibility is only considered in epistemic terms, in QM we can claim that exactly the opposite is the case, namely, that possibility is ontological and actuality has to be understood as providing an epistemic access to the truly ontological potential realm. Even though there are some similarities with modal interpretations, it is this aspect which might be regarded as the main difference between both approaches. While modal interpretations consider {\it quantum possibility} as merely epistemic, our approach remarks that {\it quantum possibility} cannot be understood in terms of classical possibility and must be regarded as the key element to produce an ontological account of QM \cite{RFD14}.

\section{Final Remarks}

The measurement problem highlights the serious difficulties to understand the path from a quantum superposition of many (sometimes even contradictory) terms to the observation of a single measurement outcome. in this paper we have criticized both actualist and hylomorphic models for grounding their questioning in the justification of ``common sense'' experience, leaving aside the signs which QM has placed for more than one century about the existence of a different realm to that of actuality. While the hylomorphic model has remained incapable of producing a metaphysical exposition of the realm of possibilities it supposedly discusses about, the actualist model has been unable to account for the physical meaning of quantum superpositions and the interaction of possibilities. 

Our new proposed model combines some of the elements found in both collapse and non-collapse models, now read from a different perspective. On the one hand, like in collapse models, we recover the realm of potentiality, but unlike them, we consider potentiality as a mode of existence completely independent to actuality leaving aside the {\it final cause} and introducing in its place the {\it immanent cause}. On the other hand, like in non-collapse solutions, we do not conceive {\bf PP} as an actual physical interaction which destroys all terms in the superposition except the one observed; rather we understand {\bf PP} as an {\it expression}, articulated through the immanent cause, of the potential realm within actuality. This understanding of powers provides an intuitive understanding and representation which is not present in the previous models of the quantum measurement process. 

Breaking down the causal teleological relation between the potential and the actual realms means to place the potential in a completely different independent ontological ground. This does not mean the elimination of the actual realm, but rather, a different and more restricted understanding of it within QM. Our proposed model offers us a new original path to investigate quantum superpositions and try to understand the relation between the quantum mechanical formalism and experience.

\begin{center}
\section*{ {\sc Appendix}}
\end{center}

\noindent Our physical representation of QM can be condensed in the following seven postulates which contain the relation between our proposed concepts and the orthodox formalism of the theory. \\

\begin{enumerate}

{\bf \item[I.] Hilbert Space:} QM is mathematically represented in a vector Hilbert space.

{\bf \item[II.] Potential State of Affairs (PSA):} A specific vector $\Psi$ with no given mathematical representation (basis) in Hilbert space represents a PSA; i.e., the definite potential existence of a multiplicity of {\it immanent powers}, each one of them with a specific {\it potentia}.

{\bf \item[III.] Quantum Situations, Immanent Powers and Potentia:} Given a PSA, $\Psi$, and the context or basis, we call a quantum situation to any superposition of one or more than one power. In general given the basis $B= \{ | \alpha_i \rangle \}$ the quantum situation $QS_{\Psi, B}$ is represented by the following superposition of immanent powers:
\begin{equation}
c_{1} | \alpha_{1} \rangle + c_{2} | \alpha_{2} \rangle + ... + c_{n} | \alpha_{n} \rangle
\end{equation}

\noindent We write the quantum situation of the PSA, $\Psi$, in the context $B$ in terms of the order pair given by the elements of the basis and the coordinates in square modulus of the PSA in that basis:
\begin{equation}
QS_{\Psi, B} = (| \alpha_{i} \rangle, |c_{i}|^2)
\end{equation}

\noindent The elements of the basis, $| \alpha_{i} \rangle$, are interpreted in terms of {\it powers}. The coordinates of the elements of the basis in square modulus, $|c_{i}|^2$, are interpreted as the {\it potentia} of the power $| \alpha_{i} \rangle$, respectively. Given the PSA and the context, the quantum situation, $QS_{\Psi, B}$, is univocally determined in terms of a set of powers and their respective potentia. (Notice that in contradistinction with the notion of {\it quantum state} the definition of a {\it quantum situation} is basis dependent and thus intrinsically contextual.)

{\bf \item[IV.] Elementary Process:} In QM we only observe discrete shifts of energy (quantum postulate). These discrete shifts are interpreted in terms of {\it elementary processes} which produce actual effectuations. An elementary process is the path which undertakes a power from the potential realm to its actual effectuation. This path is governed by the {\it immanent cause} which allows the power to remain potentially preexistent within the potential realm independently of its actual effectuation. Each power $| \alpha_{i} \rangle$ is univocally related to an elementary process represented by the projection operator $P_{\alpha_{i}} = | \alpha_{i} \rangle \langle \alpha_{i} |$.

{\bf \item[V.] Actual Effectuation of an Immanent Power (Measurement):} Immanent powers exist in the mode of being of ontological potentiality. An {\it actual effectuation} is the expression of a specific power within actuality. Different actual effectuations expose the different powers of a given $QS$. In order to learn about a specific PSA (constituted by a set of powers and their potentia) we must measure repeatedly the actual effectuations of each power exposed in the laboratory. (Notice that we consider a laboratory as constituted by the set of all possible experimental arrangements that can be related to the same $\Psi$.) An actual effectuation does not change in any way the PSA. 

{\bf \item[VI.] Potentia (Born Rule):} A {\it potentia} is the intensity of an immanent power to exist (in ontological terms) in the potential realm and the possibility to express itself (in epistemic terms) in the actual realm. Given a PSA, the potentia is represented via the Born rule. The potentia $p_{i}$ of the immanent power $| \alpha_{i} \rangle$ in the specific PSA, $\Psi$, is given by:
\begin{equation}
Potentia \ (| \alpha_{i} \rangle, \Psi) = \langle \Psi | P_{\alpha_{i}} | \Psi \rangle = Tr[P_{ \Psi} P_{\alpha_{i}}]
\end{equation}

\noindent In order to learn about a $QS$ we must observe not only its powers (which are exposed in actuality through actual effectuations) but we must also measure the potentia of each respective power. In order to measure the potentia of each power we need to expose the $QS$ statistically through a repeated series of observations. The potentia, given by the Born rule, coincides with the probability frequency of repeated measurements when the number of observations goes to infinity.

{\bf \item[VII.]  Potential Effectuations of Immanent Powers (Schr\"odinger Evolution):} Given a PSA, $\Psi$, powers and potentia evolve deterministically, independently of actual effectuations, producing {\it potential effectuations} according to the following unitary transformation:
\begin{equation}
i \hbar \frac{d}{dt} | \Psi (t) \rangle = H | \Psi (t) \rangle
\end{equation}
\noindent While {\it potential effectuations} evolve according to the Schr\"odinger equation, {\it actual effectuations} are particular expressions of each power (that constitutes the PSA, $\Psi$) in the actual realm. The ratio of such expressions in actuality is determined by the potentia of each power.
\end{enumerate}

\section*{Acknowledgements} 

I want to thank two anonymous reviewers for their careful reading of my manuscript. Their many insightful comments and suggestions allowed me to improve the article substantially. This work was partially supported by the following grants: FWO project G.0405.08 and FWO-research community W0.030.06. CONICET RES. 4541-12 and the Project PIO-CONICET-UNAJ (15520150100008CO) ``Quantum Superpositions in Quantum Information Processing''.


\end{document}